\documentstyle[aps,prl,epsfig]{revtex}

\begin{document}

\title{Lam\'e Instantons}

\author{Gerald V. Dunne and Kumar Rao}
\address{Department of Physics, University of Connecticut, Storrs CT
06269, USA}

\date{\today}
\maketitle

\begin{abstract}
We perform a precise analytic test of the instanton approximation by comparing the exact band spectrum of the periodic Lam\'e potential to the tight-binding, instanton and WKB approximations. The instanton result gives the correct leading behavior in the semiclassical limit, while the tight-binding approximation does even better. WKB is off by an overall factor of $\sqrt{e/\pi}$.
\end{abstract}
\vskip .5cm

Periodic quantum systems arise in many areas of physics, from crystal structures in solid state physics, to optical lattices in AMO physics, to solitons in polymer physics, and to the vacuum structure of QCD. An important general feature of these systems is the phenomenon of quantum tunneling that broadens discrete energy spectra into bands. Various semiclassical techniques have been developed to analyze the spectra of such systems: the tight-binding, WKB, and instanton approximations. In quantum field theory, the instanton approach provides important insights into symmetry breaking and the QCD vacuum \cite{qcd}. In this paper we compare these approximations for an exactly solvable periodic system -- the Lam\'e model -- with particular emphasis on the instanton approximation for the width of the lowest energy band. Two well-studied quantum mechanical instanton models \cite{coleman,rajaraman,shuryak} are the double-well, $V(\phi)=\frac{1}{2}\phi^2(1-\sqrt{g}\phi)^2$, and the `Sine-Gordon',  $V(\phi)=\frac{1}{g}(1-\cos(\sqrt{g}\phi))$. The Lam\'e model, $V(\phi)=\frac{1}{2g}\, {\rm sn}^2(\sqrt{g}\, \phi|\nu)$, provides a new example, and has the advantage that it can be solved exactly without the instanton approximation, while the instanton calculation can also be done analytically. We find that there is a difference between the semiclassical and tight-binding limits, limits that are often regarded as synonymous. We also highlight a new connection between the non-perturbative instanton approach and the algebraic approach to spectra \cite{franco,kusnezov}. 

Consider the Schr\"odinger-like Lam\'e equation 
\begin{eqnarray}
-\frac{d^2}{d\phi^2} \Psi(\phi)+j\, (j+1)\nu\, {\rm sn}^2(\phi |\nu) \Psi(\phi) = {\cal E} \Psi (\phi)
\label{lame}
\end{eqnarray}
where $j=1,2,3,\dots$ is an integer, ${\rm sn}(\phi|\nu)$ is one of the Jacobi elliptic functions, and $0\leq\nu\leq 1$ is the elliptic parameter. This  equation is known to be exactly solvable in terms of theta functions, and is known to have a spectrum of $j$ bound bands \cite{ww}. The function 
${\rm sn}^2(\phi|\nu)$ is periodic, with period $2K(\nu)$, where $K(\nu)=\int_0^{\pi/2}d\theta /\sqrt{1-\nu \sin^2 \theta}$ is the elliptic quarter period. For small $\nu$ (say $\nu < 0.2$), ${\rm sn}^2(\phi|\nu)$ looks to the eye like $\sin^2(\phi)$. For large $\nu$ (say $\nu > 0.9$), ${\rm sn}^2(\phi|\nu)$ looks like a sequence of periodically displaced P\"oschl-Teller $-{\rm sech}^2(\phi)$ potentials (see Fig. 1), displaced with a period that diverges logarithmically, $2 K(\nu)\sim \log(\frac{16}{1-\nu})$, as $\nu\to 1$. 

\begin{figure}[htb]
\vskip -1cm
\centering{\epsfig{file=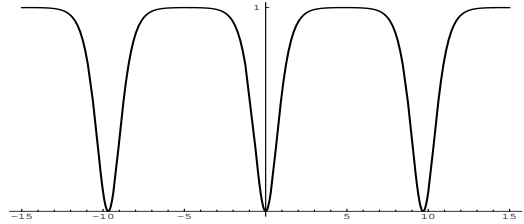, width = 4in, height=3.5in}}
\vskip -4.5cm
\caption{The function ${\rm sn}^2(\phi|\nu)$, with $\nu=0.999$, as a function of $\phi$. The period is $2K(0.999)=9.68$.}
\end{figure}

While the Lam\'e equation is exactly solvable for any $j$ and $\nu$, explicit expressions for the wavefunctions and energies are cumbersome for $j\geq 2$ \cite{ww}. A dramatic simplification is that the band-edge energies of (\ref{lame}) are given by the $2j+1$ eigenvalues of the {\it finite dimensional} matrix $J_x^2+\nu J_y^2$, where $J_x$ and $J_y$ are $su(2)$ generators in a spin $j$ representation \cite{franco,ward}. Thus we can evaluate the width $\Delta{\cal E}$ of the lowest band as the difference between the two smallest eigenvalues of the matrix $J_x^2+\nu J_y^2$. For example, for $j=1$, $\Delta{\cal E}^{\rm exact}=1-\nu$; for $j=2$, $\Delta{\cal E}^{\rm exact}=-1-\nu+2\sqrt{\nu^2-\nu+1}$; and for $j=3$, $\Delta{\cal E}^{\rm exact}=3(1-\nu)+2\sqrt{1-\nu+4\nu^2}-2 \sqrt{4-\nu+\nu^2}$. 
We will be interested later in large values of $j$, in which case such explicit expressions, as functions of $\nu$, become more difficult to derive. Instead, we have shown algebraically that for any $j$, the exact energy splitting of the lowest band, as $\nu\to 1$, is
\begin{eqnarray}
\Delta{\cal E}^{\rm exact} = {8j\, \Gamma(j+1/2)\over 4^j \sqrt{\pi}\, \Gamma(j)} \, (1-\nu)^j\, \left(1+\frac{j-1}{2} (1-\nu)+\dots\right)
\label{exact}
\end{eqnarray}
As $\nu\to 1$ the period becomes infinite, which suppresses tunneling, and we therefore expect this to be relevant for the tight-binding and semiclassical approximations. Note that the band-width (\ref{exact}) vanishes as $\nu\to 1$, and vanishes more rapidly for larger values of $j$. We now compare this exact result (\ref{exact}) with the tight-binding, instanton and WKB approximations. 

In the tight-binding approximation of solid state physics \cite{peierls} the separation $L$ between neighboring wells becomes large and we treat the potential as a sum of periodically displaced `atomic' potential wells: $V(\phi)=\sum_n U(\phi- n L)$. In the Lam\'e case, this physical approximation is explicitly realized by the remarkable mathematical identity \cite{id}:
\begin{eqnarray}
\nu\, {\rm sn}^2(\phi |\nu) =\frac{E^\prime(\nu)}{K^\prime(\nu)}-(\frac{\pi}{2K^\prime(\nu)})^2
\sum_{n=-\infty}^\infty {\rm sech}^2\left(\frac{\pi}{2K^\prime(\nu)}(\phi-2 n K(\nu))\right)
\label{magic}
\end{eqnarray}
As expected, each `atomic well' has the form of a P\"oschl-Teller well (but the rescaling factor $\frac{\pi}{2K^\prime}$ is non-obvious). Including the  $j(j+1)$ factor from (\ref{lame}), each atomic well has $j$ discrete bound states, and the effect of the periodic sum is to broaden these states into the $j$ bound bands of the Lam\'e potential. For the lowest band we use the ground state $\Psi_0(\phi)=\sqrt{\frac{\sqrt{\pi}\Gamma(j+1/2)}{2K^\prime\Gamma(j)}}  
{\rm sech}^j(\frac{\pi}{2K^\prime}\phi)$ of the atomic well. The width of this band can be calculated using standard solid state techniques \cite{peierls}:
\begin{eqnarray}
\Delta{\cal E}^{\rm tight-binding}&=& 4 j(j+1) (\frac{\pi}{2K^\prime})^2 \int_{-\infty}^\infty d\phi\, \sum_{n\neq 0}{\rm sech}^2 \left(\frac{\pi}{2K^\prime}(\phi-2 n K)\right) \Psi_0(\phi) \Psi_0(\phi-2 K)\nonumber\\
&\approx & {8j\, \Gamma(j+1/2)\over 4^j \sqrt{\pi}\, \Gamma(j)} \, (1-\nu)^j\, \left(1+\frac{j-1}{2} (1-\nu)+\dots\right)
\label{tb}
\end{eqnarray}
where in the second line we have kept dominant terms as $\nu\to 1$, and used the fact that $\exp[-\pi K(\nu)/K^\prime(\nu)]\sim \frac{1-\nu}{16}
(1 +\frac{1}{2}(1-\nu)+\dots)$. This result (\ref{tb}) agrees precisely with the exact result (\ref{exact}) to this order in $1-\nu$. Therefore, for any $j$, the tight-binding approximation is good as $\nu\to 1$; {\it i.e.} as the separation between atomic wells becomes large.

We now turn to an instanton evaluation of the width of the lowest band, expected to be good in the semiclassical limit in which tunneling effects are small. Naively, one might expect that this is also just the $\nu\to 1$ limit in which the wells become infinitely separated, but it is actually more interesting than this. To make contact with the standard instanton approach \cite{coleman,rajaraman,shuryak} we define
\begin{eqnarray}
V(\phi)=\frac{1}{2g}\, {\rm sn}^2(\sqrt{g}\, \phi|\nu)
= \frac{1}{2}\phi^2-g\,\frac{(\nu+1)}{6}\phi^4+\dots
\label{instpot}
\end{eqnarray}
where $g$ is some coupling (which we will relate to $j$ and $\nu$ below), and we have chosen units so that the perturbative mass in a given well is $1$. The width of the lowest band can be found by considering the Euclidean path integral connecting two neighboring classical minima of the potential; here $\phi=0$ and $\phi=2K/\sqrt{g}$ for example. Rescaling the field variable to $\chi=\sqrt{g}\phi$,
\begin{eqnarray}
\exp\left(-\frac{1}{\hbar}S[\phi]\right)= \exp\left( -\frac{1}{\hbar g} \int dt\, [\frac{1}{2}(\frac{d\chi}{dt})^2+\frac{1}{2}{\rm sn}^2(\chi|\nu)]\right)
\label{semi}
\end{eqnarray}
Thus the semiclassical limit is $\hbar g\ll 1$, or $\frac{1}{2}\hbar\ll V_{\rm peak}$ : the ground state in each well is far below the barrier height. 

There is a standard technique for computing the instanton approximation for the width of the lowest band \cite{coleman,rajaraman,shuryak}. In the semiclassical limit, the Euclidean path integral is dominated by `instanton' solutions to the Euclidean equations of motion, satisfying the first-order equation $\dot{\chi}_{\rm inst}=\sqrt{2 V(\chi_{\rm inst})}$. Using the rescaled potential $V(\chi)=\frac{1}{2}{\rm sn}^2(\chi|\nu)$, we find the Lam\'e instanton:
\begin{eqnarray}
\chi_{\rm inst}(t)=K(\nu)+{\rm sn}^{-1}({\rm tanh}(t)|\nu)
\label{inst}
\end{eqnarray}
Here the integration constant has been chosen so that the instanton is centered at $t=0$, and ${\rm sn}^{-1}$ means the inverse function (a standard function in Mathematica). This instanton interpolates between $\chi=0$ at $t=-\infty$, and $\chi=2 K(\nu)$ at $t=+\infty$, as shown in Fig. 2. The corresponding Euclidean action is
\begin{eqnarray}
S_0=\int_0^{2K}d\chi  \sqrt{2 V(\chi)} =\frac{1}{\sqrt{\nu}} \log\left(\frac{1+\sqrt{\nu}}{1-\sqrt{\nu}}\right)
\label{action}
\end{eqnarray}
\begin{figure}[htb]
\vskip -1cm
\centering{\epsfig{file=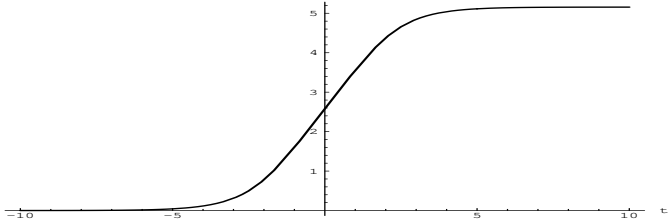, width = 5in, height=3.5in}}
\vskip -5cm
\caption{The Lam\'e instanton $\chi_{\rm inst}(t)$ in (\protect{\ref{inst}}), plotted for $\nu=0.9$. Note that $2 K(0.9)=5.15618$.}
\end{figure}
The leading exponential factor in the instanton expression for the band width is $\exp[-S_0/(\hbar g)]$. But there is also a prefactor that is related to the determinant of the fluctuation operator $-\frac{d^2}{dt^2}+V^{\prime\prime}(\chi_{\rm inst}(t))$. This prefactor is physically significant, as it encapsulates the collective coordinate effects of fluctuations about the instantons. Here,
\begin{eqnarray}
V^{\prime\prime}(\chi_{\rm inst}(t))= (1-\nu)\left[{\nu\, 
{\rm tanh}^4(t)+2(1-\nu)\,{\rm tanh}^2(t)-1\over (1-\nu\, 
{\rm tanh}^2(t))^2}\right]
\label{fluc}
\end{eqnarray}
which is plotted in Fig. 3. We see that $V^{\prime\prime}(\chi_{\rm inst}(t))$ is highly localized in the vicinity of the instanton, and tends to $1$ (the square of the mass) as $t\to \pm\infty$, as expected.
\begin{figure}[htb]
\vskip -1cm
\centering{\epsfig{file=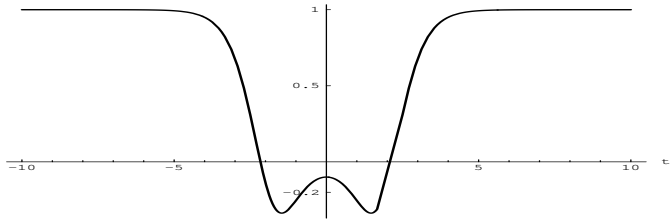, width = 5in, height=3.5in}}
\vskip -4.5cm
\caption{The fluctuation potential $V^{\prime\prime}(\chi_{\rm inst}(t))$ in (\protect{\ref{fluc}}), plotted for $\nu=0.9$.}
\end{figure}
Coleman has given a very simple method \cite{coleman} for evaluating the fluctuation determinant (with zero mode removed), resulting in the formula (for a periodic potential)
\begin{eqnarray}
\Delta E^{\rm inst}=c\, 4 \hbar \frac{1}{\sqrt{\pi \hbar g}} \, \exp[-\frac{1}{\hbar g} S_0]
\label{form}
\end{eqnarray}
where the constant $c$ is simply determined by the asymptotic behavior of the zero mode: $\dot{\chi}_{\rm inst}(t)\sim c \, e^{-|t|}$ as $|t|\to \infty$. Here, the zero mode is $\dot{\chi}_{\rm inst}(t)={\rm sech}(t)/\sqrt{1-\nu\, 
{\rm tanh}^2(t)} \sim \frac{2}{\sqrt{1-\nu}} e^{-|t|}$. Thus we obtain the Lam\'e instanton result \cite{sgcomment}
\begin{eqnarray}
\Delta E^{\rm inst}= \frac{8 \hbar}{\sqrt{\pi \hbar g}} \, \frac{1}{\sqrt{1-\nu}}\, \exp[-\frac{1}{\hbar g\sqrt{\nu}} \log\left(\frac{1+\sqrt{\nu}}{1-\sqrt{\nu}}\right)]
\label{instanton}
\end{eqnarray}
To relate this to the eigenvalues ${\cal E}$ of the Lam\'e equation (\ref{lame}), we compare (\ref{lame}) with the Schr\"odinger equation, $-\frac{\hbar^2}{2}\frac{d^2}{d\phi^2}\Psi+\frac{1}{2g} {\rm sn}^2(\sqrt{g}\phi)\Psi=E\Psi$, for the potential $V(\phi)$ in (\ref{instpot}). We therefore identify
\begin{eqnarray}
\nu\, j\,(j+1)=\frac{1}{\hbar^2 g^2}, \qquad \qquad {\cal E}=
\frac{2 E}{\hbar^2 g}
\label{corr}
\end{eqnarray}
Thus the instanton approximation for the band-width of the eigenvalue 
${\cal E}$ in (\ref{lame}) is
\begin{eqnarray}
\Delta {\cal E}^{\rm inst}=\frac{16}{\sqrt{\pi}} \left(\nu j\,(j+1)\right)^{3/4} \, \left(1+\sqrt{\nu}\right)^{-2\sqrt{j(j+1)}}\, \left(1-\nu\right)^{\sqrt{j(j+1)}-1/2} 
\label{final}
\end{eqnarray}
As a function of $j$ this differs from the exact and tight-binding expressions (\ref{exact}) and (\ref{tb}), even in the large period limit $\nu\to 1$. However, from (\ref{semi}) and (\ref{corr}), the semiclassical limit $\hbar g\ll 1$ means $\nu j(j+1)\gg 1$. So, to compare the instanton formula (\ref{final}) with the results (\ref{exact}) and (\ref{tb}) it is not enough to take $\nu\to 1$; we also need to take $j$ to be large. Physically, it is not enough to take far-separated wells; they must also be deep wells. For large $j$ and $\nu$ near 1,
\begin{eqnarray}
\Delta {\cal E}^{\rm inst}\sim \frac{8 j^{3/2}}{\sqrt{\pi}\, 4^j} \, \left(1-\nu\right)^j \left(1+\frac{j-1}{2}(1-\nu)+\dots \right)
\label{largej}
\end{eqnarray}
which agrees perfectly with the large $j$ limit (using Stirling's formula) of the exact result (\ref{exact}). 

As another test of our instanton formula (\ref{final}) we can fix $\nu$ to any value (not necessarily near 1) and take $j$ large (in order to be in the semiclassical regime), and compare to the band-width obtained numerically from the two lowest eigenvalues of the matrix $J_x^2+\nu J_y^2$. The results are shown in Fig. 4, showing $10 \% $ agreement for $j\geq 10$. In terms of the algebraic spectral program \cite{franco,kusnezov}, we find that non-perturbative instantons play an interesting role in the semiclassical limit via eigenvalue differences for finite dimensional, but very large matrices. As a final calculation, we have computed the WKB approximation for the energy splitting, and we find that in the semiclassical limit $\Delta E^{\rm WKB}=\frac{2\hbar}{\pi} 
\exp[-\frac{1}{\hbar g}\int_{\rm TP} d \chi 
\sqrt{{\rm sn}^2(\chi)-\hbar g}] =\sqrt{\frac{e}{\pi}} \Delta E^{\rm inst}$, confirming that WKB gets the correct leading exponential but the WKB prefactor normalization is incorrect \cite{gild}. 
\begin{figure}[htb]
\vskip -1cm
\centering{\epsfig{file=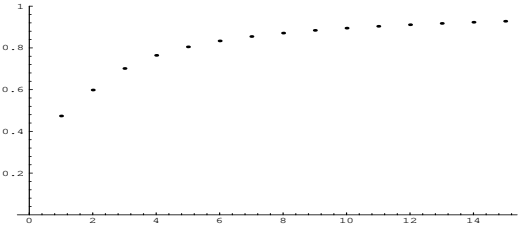, width = 4in, height=3.5in}}
\vskip -4.5cm
\caption{The ratio of the exact numerical band-width to the instanton expression (\protect{\ref{final}}), for various values of $j$, with $\nu=0.5$.}
\end{figure}
To conclude, the instanton approximation gives the correct leading semiclassical result for the lowest band-width. But the Lam\'e model (\ref{lame}) has two independent parameters, $\nu$ and $j$, that allow us to probe separately the `far-separated-well' and `deep-well' limits respectively. The tight-binding approximation (something of a misnomer) is good for large period, for any well-depth; while the instanton approximation requires deep wells for a given period so that the combination $\nu j(j+1)$ is large. It would be interesting to study instanton - anti-instanton interactions and correlation functions in the Lam\'e system, and to compare to lattice simulations in the spirit of \cite{shuryak2}. Also, Lam\'e solitons with spatial profile of the form (\ref{inst}) will arise in the corresponding $1+1$ dimensional model.

This work was supported in part by DOE grant DE-FG02-92ER40716.00.


\begin{references}
\bibitem{id} Here $K^\prime(\nu)=K(1-\nu)$, $E^\prime(\nu)=E(1-\nu)$, and $E(\nu)$ is the complete elliptic integral \cite{ww}.
\bibitem{sgcomment} The Lam\'e model is a deformation of the Sine-Gordon case, as setting $\nu=0$ [instead of making the identification in (\protect{\ref{corr}})] in formulas (\protect{\ref{instpot}} - \protect{\ref{instanton}}) reproduces the well-known instanton results for Sine-Gordon. This is because ${\rm sn}^2(\phi|\nu=0)\equiv\sin^2(\phi)$.
\end{references}
\end{document}